The GoogLeNet-assisted phase transition detectors

[1]C.H.Wong*, [2]Raymond P. H. Wu, [1]X.Lei, [1]A.F.Zatsepin

[1]Institute of Physics and Technology, Ural Federal University, Russia

[2]Department of Physics, The Chinese University of Hong Kong, Hong Kong

*ch.kh.vong@urfu.ru

**Abstract:**

In the presence of the phase fluctuations in superconducting nanowires array, the electrical resistance of the superconducting nanowires is always non-zero unless the system undergoes Berezinskii-Kosterlitz-Thouless (BKT) transition where the superconducting vortices and anti-vortices form pairs. The two-dimensional XY model can mimic the superconducting transition temperature $T_c$ and the BKT transition at a lower critical temperature $T_{BKT}$ by observing the heat capacity anomalies upon cooling. If the Josephson coupling across the nanowires is strong, the heat capacity anomalies almost overlap with each other so that it is difficult to distinguish between the $T_c$ and the $T_{BKT}$. To solve this issue, we apply an artificial-intelligence technique to split the nearly overlapped heat capacity anomalies. After the GoogLeNet-assisted phase transition detector is built, the GoogLeNet model can learn from the features of the phase transitions and then interpret the $T_c$ and $T_{BKT}$ in the 'unseen' system precisely. Our work opens a path for the GoogLeNet model to enter the world of magnetism and superconductivity.

**Introduction:**

To vanish the electrical resistance of superconductors, the phase and amplitude of the superconducting parameters must be ordered [1,2]. Unlike bulk superconductors, the phase fluctuations in low-dimensional superconductors suppress the formation of a long-range ordered state even if the temperature is below the $T_c$ [1,2]. In superconducting films, the phase stability can be achieved by undergoing the Berezinski–Kosterlitz–Thouless (BKT) transition [3,4,5] at a lower critical temperature $T_{BKT}$, where a quasi-long range superconducting order is created. Despite the van Hove singularities [6] in the electronic density of states of 1D superconductors would be advantageous to produce a high $T_c$ phenomenon, the superconducting phase slips at 1D limit suppressing a long-range order at any finite temperature associated with a finite electrical resistance [7-11]. However, the disappearance of the phase fluctuations has been observed in the quasi-1D superconducting nanowires array under a transverse Josephson coupling where a long-range-ordered superconducting state with zero resistance at finite temperature is obtainable [12-19].

In recent years, low temperature physicists aim at raising the $T_{BKT}$ closer to the $T_c$ [12,14,15]. The Monte Carlo XY model can be used to mimic the BKT transition and the superconducting phase transition [4,20]. The critical temperautures for the system with the Josephson coupling J can be determined by searching the locations of the heat capacity anomalies [4]. The heat capacity anomaly at a higher critical temperature ($T_c$) refers to the formation of the Cooper pairs [1,2]. On the other hand, the heat capacity anomaly at a lower critical temperature ($T_{BKT}$) corresponds to the formation of the vortex - antivortex pairs [4]. If the Josephson coupling is weak [14], these two heat capacity anomalies are widely separated at different temperatures [13]. Otherwise, the heat capacity anomalies at the $T_c$ and the $T_{BKT}$ are almost overlapped [12,15] and hence it is difficult to identify the phase transition temperatures accurately.

To solve this issue, a phase transition detector without showing multiple phase transition temperatures on the temperature domain is needed. We propose to combine the 8-state XY model (q = 8) and the GoogLeNet model to detect the $T_c$ and the $T_{BKT}$. The GoogLeNet designed by the research teams at Google with the collaboration of several universities has been a forefront computer vision algorithm since 2014. [21] The GoogLeNet model contains 1×1 convolutions in the middle of the architecture and global average pooling that establishes a deeper architecture of 22 layers deep with low computational costs [22]. With an excellent architecture in the GoogLeNet model, we test whether or not the GoogLeNet-assisted phase transition detector can show the $T_c$ and the $T_{BKT}$ separately.

**Methodology:**

The Monte Carlo method with Metropolis algorithm is used to simulate the standard 8-state XY model (q = 8) on a square lattice, where the electron-electron interaction J is spatial independent along the sample surface [4,7,12]. The phase-maps as a series of temperature at equilibrium state is obtained. Then the phase-maps are converted in JPEG format for image classification.

To identify the $T_c$ at which two electrons form a Cooper pair (CP) [7], two thousand phase-maps with the size of 66 x 66 are simulated at the dimensionless $T_c$ under the condition of J = 1, which is named as the "stock data (CP)". To locate the $T_{BKT}$ at which the vortex and anti-vortex form a pair [4,20], two thousand phase-maps with the size of 66 x 66 are generated at the dimensionless $T_{BKT}$ under the condition of J = 1, which is named as the "stock data (BKT)". Then we create the "test data (q = 8, J = 1)" and the "test data (q = 8, J = 0.8)" where two thousand 66 x 66 phase-maps are generated at each temperature, respectively.

The image classification is based on the GoogLeNet model under the Convolutional Neural Network (CNN) [21,22]. The image database is split into training data (70%) and validation data (30%). If the GoogLeNet model can recognize the phase-maps accurately, the validation accuracy of phase-maps is close to 100%. To check whether the sample is located at the phase transition temperatures or not, the likelihood of the phase transitions, i.e. LH factor (CP) and LH factor (BKT), are defined as (100% - validation accuracy) / 50%. When the LH factor closes to 1, the material is expected to undergo the phase transition.

**Results and Discussions:**

Figure 1 shows the thermal fluctuations which are generated by the 8-state XY model with J = 1 and J = 0.8. After a careful curve-fitting process, we get the $T_{BKT}$ = 0.34 and the $T_c$ = 1.13 for J = 1 in Figure 1a. Similarly, we get the $T_{BKT}$ = 0.32 and the $T_c$ = 0.85 for J = 0.8 in Figure 1b. This verify that the values of $T_{BKT}$ and $T_c$ depend on the electron-electron interaction J. The two thousand phase-maps simulated in the conditions of J = 1 and $T_c$ = 1.13 are grouped as the "stock data (CP)". Meanwhile, the two thousand phase-maps in the "stock data (BKT)" are created in the conditions of J = 1 and $T_{BKT}$ = 0.34. The phase-maps are plotted in Figure 2, respectively. The direction of phase is displayed in terms of color-plot which is advantageous to image classification. In the 8-state XY model, the direction of superconducting phase at 0, 45, 90, 135, 180, 225, 270, 315 degrees refers to the integer 0, 1, 2, 3, 4, 5, 6, 7 in the color bar, respectively.

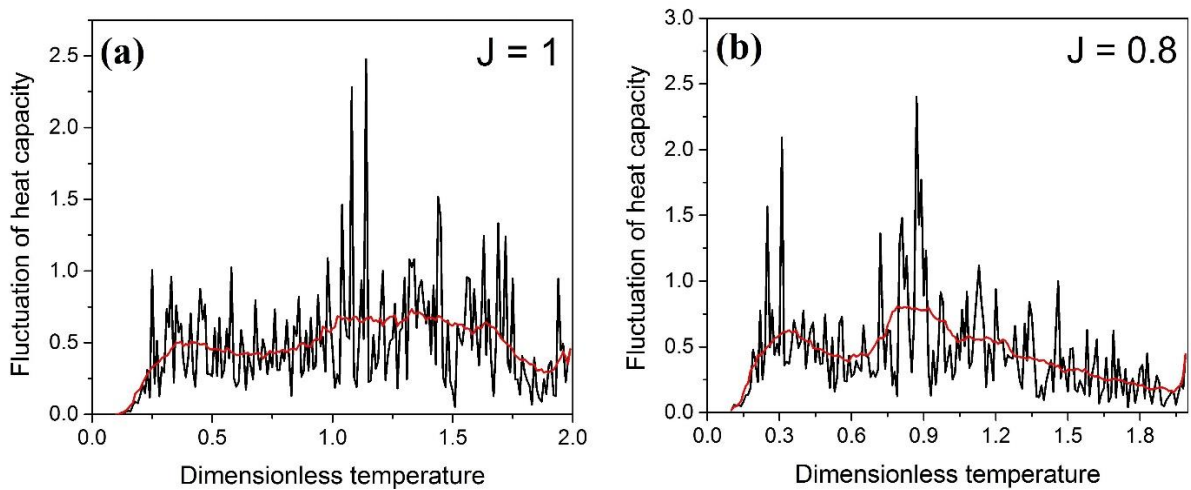

Figure 1: The fluctuations of heat capacity as a function of temperature (a) J = 1. (b) J = 0.8.

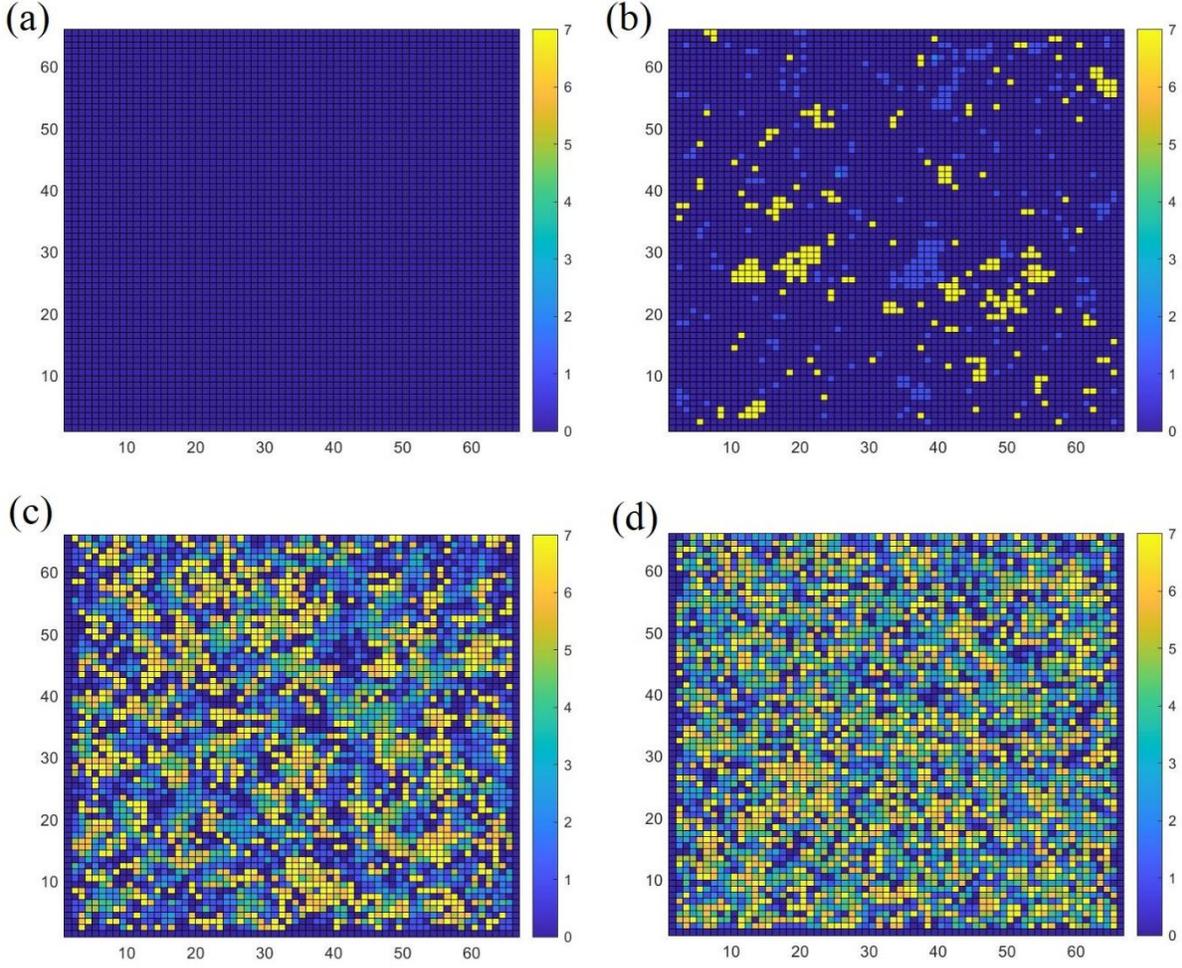

Figure 2: The phase-maps for J = 1 at different temperatures. (a) The phase-map below the $T_{BKT}$. (b) The phase-map at the $T_{BKT}$. (c) The phase-map at the $T_c$. (d) The phase-map above the $T_c$.

According to Figure 3, the GoogLeNet-assisted phase transition detector avoids showing the $T_c$ and the $T_{BKT}$ on the same domain successfully. Although the traditional method [4] in Figure 1a shows that the BKT transition occurs at the $T_{BKT} \sim 0.3$, Figure 3a shows that the LH factor (CP) remains zero for T < 0.5. Zero LH factor (BKT or CP) means that undergoing the phase transition is not possible. For J = 1, the GoogLeNet-assisted phase transition detector shows that the $T_c$ is 1.1 with the evidence of the LH factor (CP) = 1. On the other hand, replacing the "stock data (CP)" by the "stock data (BKT)" emerges the BKT transition at $T_{BKT} \sim 0.3$ as shown in Figure 3b, where the LH factor (BKT) = 1. The LH factor (BKT) returns to zero for T > 0.6 even though the superconducting phase transition occurs at $T_c \sim 1$.

As the stock data and the test data in Figure 3 are computed by setting the J = 1 in the XY model [4], we proceed to examine the accuracy of the GoogLeNet-assisted phase transition detector if the electron-electron interaction in the stock data and the test data are not identical. Figure 4 demonstrates that the GoogLeNet model [21,22] is capable of learning the features from the phase-maps in the J = 1 system and then interprets the unknown phase transition temperatures in the 'unseen' systems (J ≠ 1).

The prediction of the GoogLeNet model in the J = 0.8 system is also accurate. Although Figure 1b illustrates the BKT transition at T ~ 0.3, Figure 4 confirms that the GoogLeNet-assisted phase transition detector can identify the $T_c$ and the $T_{BKT}$ separately. After the J is reduced to 0.8 in the "test data", Figure 4a shows that the LH factor (CP) is the highest at $T_c$ = 0.83 where the superconducting phase transition takes place. No BKT transition is observed in Figure 4a. Again, in Figure 4b, changing the "stock data (CP)" to the "stock data (BKT)" emerges the BKT transition at T ~ 0.3 and the superconducting phase transition at T ~ 1 is disappeared. The LH factor (BKT) becomes zero for T > 0.7 in Figure 4b.

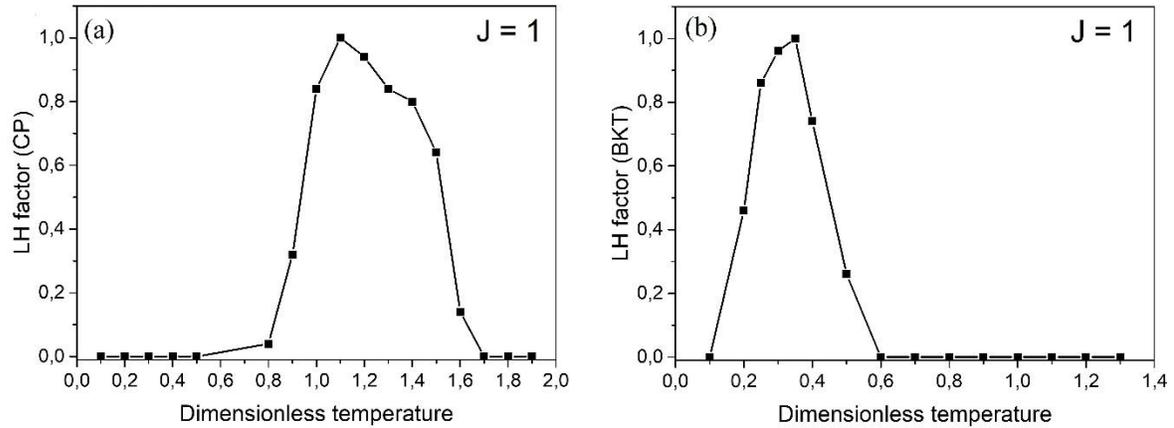

Figure 3: The LH factors as a function of temperatures. The stock data (CP), the stock data (BKT) and the test data are run by assigning the J = 1 in the XY model (a) The LH factor (CP) refers to the likelihood of forming the superconducting phase transition. (b) The LH factor (BKT) refers to the likelihood of undergoing the BKT transition.

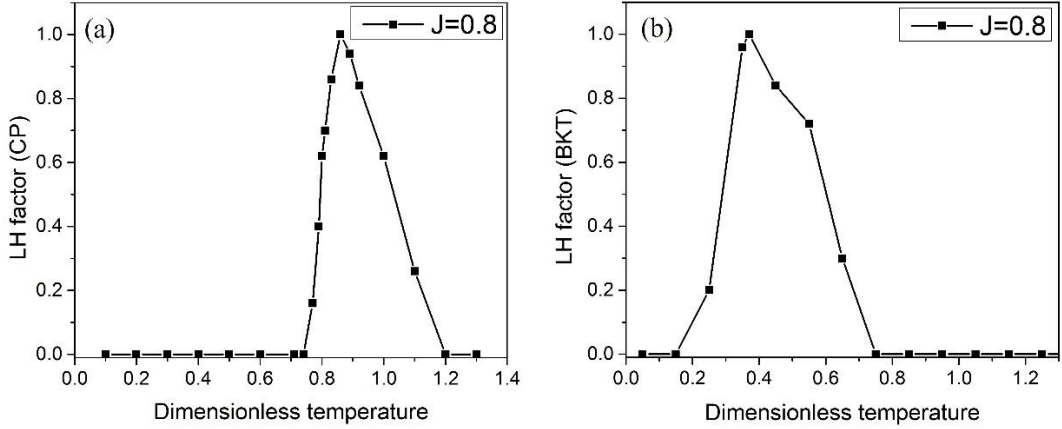

Figure 4: The LH factors as a series of temperatures. The stock data (CP), the stock data (BKT) are run by assigning the J = 1 in the XY model. The test data are prepared by setting the J = 0.8 in the XY model. (a) The LH factor (CP) of 1 corresponds to the superconducting phase transition temperature. (b) The LH factor (BKT) of 1 refers to the BKT transition temperature.

When we draw parallel between Figure 1, Figure 3 and Figure 4, the phase transition temperatures probed by the GoogLeNet model [21,22] are almost identical to the traditional method of heat capacity measurement [4,12,20]. However, there is a big advantage of using the GoogLeNet-assisted phase transition detector: the superconducting phase transition and the BKT transition can be observed separately. If the traditional method is used to probe the phase transition temperature [4], these two heat capacity anomalies emerge on the same temperature domain in Figure 1, which requires advanced numerical methods [12-15] to distinguish the $T_{BKT}$ and the $T_c$. Without the advanced curve fitting techniques, one may argue that there are three heat capacity anomalies at T ~ 0.3, T ~ 1 and T ~ 1.7 in Figure 1a. After we fit the curves carefully with help of the scientific knowledge and numerical mathematics [12-15], we can figure out that the peaks at T ~ 1.7 do not refer to the superconducting phase transition temperature. In contrast, probing the superconducting transition temperature via the use of the LH factor can avoid this problem because the non-zero LH factor (CP) and LH factor (BKT) do not appear at the same temperature.

Let us recall that a higher validation accuracy of the phase-maps produces a lower LH factor. When the temperature is near to the $T_c$ or the $T_{BKT}$, the phase-maps in the "stock data" and the "test data" look similar. Then it reduces the validation accuracy and raises the LH factor where we make use of this trend to locate the phase transition temperatures. The GoogLeNet-assisted phase transition detector is accurate as shown in Figure 3. It is reasonable because the electron-electron interactions in the "stock data" and the "test data" are the same. As the GoogLeNet model has learnt the features from the stock data at J =1, it interprets the phase-maps in the

'unseen' system (J = 0.8) successfully, because the distribution of phase at the $T_c$ (or $T_{BKT}$) shares the same nature regardless of the value of J.

We used a FUJITSU computer equipped with Intel(R) Core(TM) i7-3632QM CPU at 2.2GHz to train the phase transition detector that requires 25 minutes only. With this reasonable computational cost, it is encouraging to predict the phase transitions by using the artificial-intelligence assisted XY model.

**Conclusions:**

To avoid the two heat capacity anomalies generated by the XY model appearing on the same temperature domain, the GoogLeNet model under the Convolutional Neural Network is applied to analyze the distribution of phase in the superconducting nanowires array. With the excellent image classification of the GoogleNet model, our phase transition detector can interpret the influence of the Josephson coupling on the Berezinski–Kosterlitz–Thouless (BKT) transition and predict the effect of the pairing strength on the superconducting transition temperature.